# Optimization of Time-Resolved Magneto-optical Kerr Effect Signals for Magnetization Dynamics Measurements

Dustin M. Lattery[1], Delin Zhang[2], Jie Zhu[1], Paul Crowell[3], Jian-Ping Wang[2] and Xiaojia Wang[1]*

[1]Department of Mechanical Engineering, University of Minnesota, Minneapolis, MN 55455, USA

[2]Department of Electrical and Computer Engineering, University of Minnesota, Minneapolis, MN 55455, USA

[3]School of Physics and Astronomy, University of Minnesota, Minneapolis, MN 55455, USA

*Corresponding authors: wang4940@umn.edu

**Abstract:** Recently magnetic storage and magnetic memory have shifted towards the use of magnetic thin films with perpendicular magnetic anisotropy (PMA). Understanding the magnetic damping in these materials is crucial, but normal Ferromagnetic Resonance (FMR) measurements face some limitations. The desire to quantify the damping in materials with PMA has resulted in the adoption of Time-Resolved Magneto-optical Kerr Effect (TR-MOKE) measurements. In this paper, we discuss the angle and field dependent signals in TR-MOKE, and utilize a numerical algorithm based on the Landau-Lifshitz-Gilbert (LLG) equation to provide information on the optimal conditions to run TR-MOKE measurements.

## I. INTRODUCTION

Spintronics utilizing perpendicular magnetic anisotropy (PMA) are very promising for the advancement of computer memory, logic, and storage. Due to the time scale of magnetic switching in these devices (~ 1 ns), it is crucial to understand the ultrafast dynamic magnetization, which behave according to the Landau-Lifshitz-Gilbert (LLG) equation. The application of this equation to understand magnetization dynamics requires knowledge of the magnetic anisotropy and the Gilbert damping (α). While anisotropy can be determined through magnetostatic measurements, extracting α requires measurements that can capture the dynamic magnetization at time scales faster than magnetic switching. To date, the most common method to do this is through frequency domain measurements of ferromagnetic resonance (FMR). By measuring the resonance frequency and linewidth as a function of field, FMR can probe both the magnetic anisotropy and Gilbert damping. As spintronic applications begin to use materials with large PMA, the use of another technique, time-resolved magneto-optical Kerr effect (TR-MOKE), has increased. This technique (which is essentially a time-domain FMR measurement technique) is able to measure at higher resonance frequencies and external fields, which allows extremely hard magnetic materials to be measured.

There are many papers discussing TR-MOKE measurements for measuring the Gilbert Damping. Most of these papers utilize similar polar MOKE measurement techniques, but there is often a large variation in both the $H_{ext}$ range for measurements and in the angle of external field. While some papers utilize in-plane external field because of its well-understood frequency dependence, others choose to apply the field at a chosen angle away from the surface normal. It has been theorized and shown in measurements that the process of applying the field at some angle between 0 and 90° is beneficial to increase the TR-MOKE signal amplitude, but the explanations as to why this occurs are lacking. In this paper, we aim to discuss why the signal depends on the angle of external field and calculate the optimal angle for conducting TR-MOKE measurements of damping on magnetic materials with PMA.

## II. FINITE DIFFERENCE METHOD LANDAU-LIFSHITZ-GILBERT EQUATIONS

Simulations in this work utilize a finite difference approach to solve the LLG equation (Eq. 1) with an explicit solution for the magnetization vector (**M**) as a function of time following the forward Euler method.

$$\frac{d\mathbf{M}}{dt} = -\gamma\left(\mathbf{M}\times\mathbf{H}_{\text{eff}}\right) + \frac{\alpha}{M_s}\left(\mathbf{M}\times\frac{d\mathbf{M}}{dt}\right) \quad (1)$$

where **M** is the magnetization vector with a magnitude of $M_s$ (the saturation magnetization), $\gamma$ is the gyromagnetic ratio, $\mathbf{H}_{\text{eff}}$ is the effective magnetic field, and $\alpha$ is the Gilbert damping parameter. The vector $\mathbf{H}_{\text{eff}}$ is determined by taking the gradient of the magnetic free energy density ($F$) with respect to the magnetization direction ($\mathbf{H}_{\text{eff}} = -\nabla_{\mathbf{M}} F$). The scalar quantity $F$ is the summation of contributions from Zeeman energy (from the external magnetic field, $\mathbf{H}_{\text{ext}}$), perpendicular uniaxial magnetic anisotropy ($K_u$), and the demagnetizing field (assuming the sample is a magnetic thin film).

While Eqn. 1 is often used to describe magneto-dynamics due to the use of $\alpha$, it is not conducive to numerical solutions of this ordinary differential equation. To simplify the development of computational algorithms, it is preferential to utilize the Landau-Lifshitz equation (Eq. 2).

$$\frac{d\mathbf{M}}{dt} = -\gamma'\left(\mathbf{M}\times\mathbf{H}_{\text{eff}}\right) - \frac{\lambda}{M_s^2}\left(\mathbf{M}\times\mathbf{M}\times\mathbf{H}_{\text{eff}}\right). \quad (2)$$

The coefficients in Eq. 2 can be related to the previously defined constants in Eqs. 3 and 4 [1].

$$\gamma' = \frac{\gamma}{1+\alpha^2} \quad (3)$$

$$\lambda = \alpha\gamma' M_s \quad (4)$$

In equilibrium, **M** is parallel to **H**eff, and so the magnetization does not precess. If the magnetization is removed from the equilibrium direction, it will begin precessing around the equilibrium direction, finally damping towards equilibrium at a rate determined by the magnitude of $\alpha$ (shown in Fig. 1).

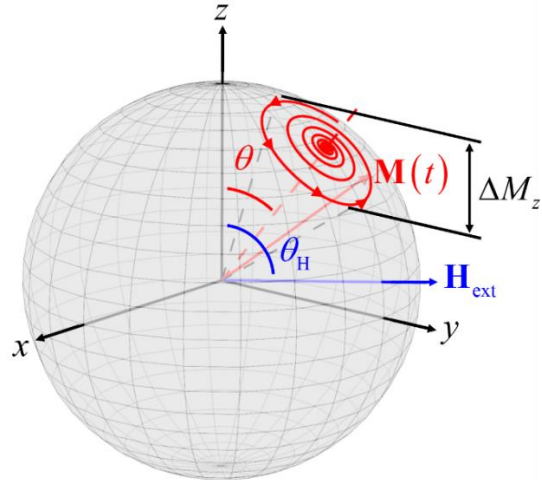

**Figure 1**. A three-dimensional representation of the magnetization vector (**M**) precessing around the equilibrium direction ($\theta$) displayed on the surface of a sphere of radius $M_s$. The equilibrium direction is controlled by the magnitude and direction ($\theta_H$) of the external magnetic field vector (**H**$_{ext}$). The change in the z-component of magnetization ($\Delta M_z$) is proportional to the TR-MOKE signal.

To initiate precession, a thermal demagnetization process is applied, emulating TR-MOKE measurements. For TR-MOKE measurements, a "pump" laser pulse increases the temperature at an ultrafast time scale, causing a thermal demagnetization (a decrease in $M_s$ caused by temperature) [2, 3]. This thermal demagnetization temporarily moves the equilibrium direction causing the magnetization to begin precession, which is continued even when $M_s$ has recovered to its original state. Here, the demagnetization process is treated as a step decrease in $M_s$ that lasts for 2.5 ps before an instant recovery to the initial value. All signal analysis discussed in this work is following the recovery of $M_s$.

For polar MOKE measurements, the projected magnetization in the z-direction ($M_z$, through-plane magnetization) is proportional to the Kerr rotation [4]. The projection of $M_z$ in time during precession will appear is a decaying sinusoid ($M_z(t) \propto \sin(\omega t + \varphi)\exp(-t/\tau)$), which is also captured by TR-MOKE measurements. The amplitude of the precession will greatly depend on the applied field magnitude and angle, which is also carried into TR-MOKE signal. By analyzing the precession as a function of field and angle, the precession amplitude (delta Mz) can be extracted. Figure 2 shows the process of extracting the amplitude as a function of angle for two different regions of magnetic field. Tracking this signal amplitude as a function of $\theta_H$, reveals that the precession (and thus the signal) will be maximized for a certain $\theta_H$ as shown in Fig. 2(b). Maximizing the oscillation implies that it will be beneficial to maximize the "magnetic torque" term ($M \times H_{eff}$, which prefers a large angle between $M$ and $H_{eff}$), but it also important to factor in that TR-MOKE measures the projection of the magnetization along the z-direction (which prefers $\theta = 90°$). Because of this, the value of $\theta_{H,MAX}$ requires weighing inputs from both the magnetic torque and the z-direction projection of magnetization.

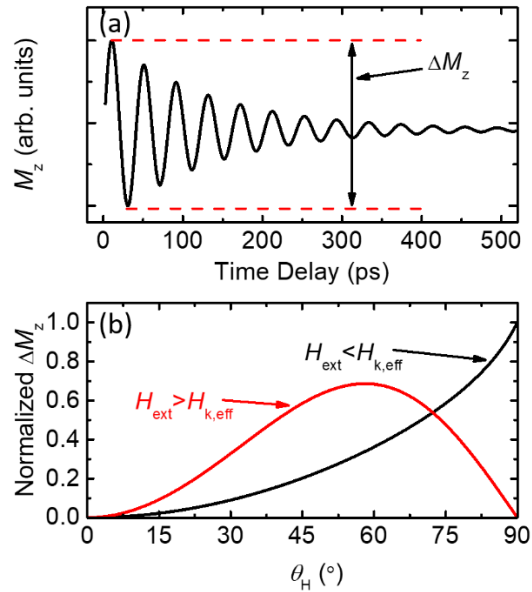

**Figure 2**. For specific conditions, the LLG simulation will produce a time-dependent magnetization vector. The difference between the maximum and minimum of the z-component of magnetization in time ($\Delta M_z$) provides information about the strength of the TR-MOKE signal. These simulations are conducted for a range of $\theta_H$ resulting in the curves in (b). The trend of signal with increasing $\theta_H$ also depends on the magnitude of the external field relative to $H_{k,eff}$, as shown by the black ($H_{k,eff} < H_{ext}$) and red ($H_{k,eff} > H_{ext}$) lines.

Depending on whether the field ratio ($H_{ext}/H_{k,eff}$) the angular dependence on magnitude will drastically change. For $H_{ext} < H_{k,eff}$, the magnetization will be in equilibrium between the perpendicular direction and the in-plane direction ($0 \leq \theta \leq 90°$). Maximizing the magnetic torque and projection in the z-direction in these cases will cause $H_{ext}$ applied in-plane ($\theta_H = 90°$) to be the optimal setup [shown by the red line in Fig. 4(b)]. Once $H_{ext}$ exceeds $H_{keff}$, the Stoner-Wolfarth minimum energy model predicts that the magnetization will approach the direction of external field, but never align (except along $\theta_H = 0$ or $90°$). Because these two directions will have no magnetic torque, there should be no magnetic precession, and thus there will be amplitude minima at these extremes. Between these two angles, the two effects for optimizing signal will complete, leading to an amplitude maximum at an angle that depends on the size of the ratio.

Figure 3 shows a contour plot of the dependence on signal amplitude as a function of both the magnitude of $H_{ext}$ and $\theta_H$. The highest amplitude of precession will occur near $H_{k,eff}$ when the field is applied in the film plane. If the external field is greater than $H_{k,eff}$, it is beneficial to conduct the measurement at an angle that is out of the film plane. To reveal this trend, the dotted red line in Fig. 3 indicates the angle of maximum signal ($\theta_{H,MAX}$) as a function at specified field ratios. Note that the curve does not follow the gradient of signal vs. $\theta_H$ and $H_{ext}$. This is due to the definition of $\theta_{H,MAX}$ as the value of $\theta_H$ that maximizes signal for a given *Hext*, instead of a maximization of signal with both parameters. Based on these results, measurement conditions can be tuned to maximize the signal based on the field ratio. For example, if the maximum strength of the magnetic field is only $2H_{k,eff}$, then it would be beneficial to set $\theta_H > 60°$. Furthermore, measurements conducted at a constant field and a varied magnetic field angle, should not necessarily conduct the measurement at the highest possible $H_{ext}$ if the goal is to maximize SNR.

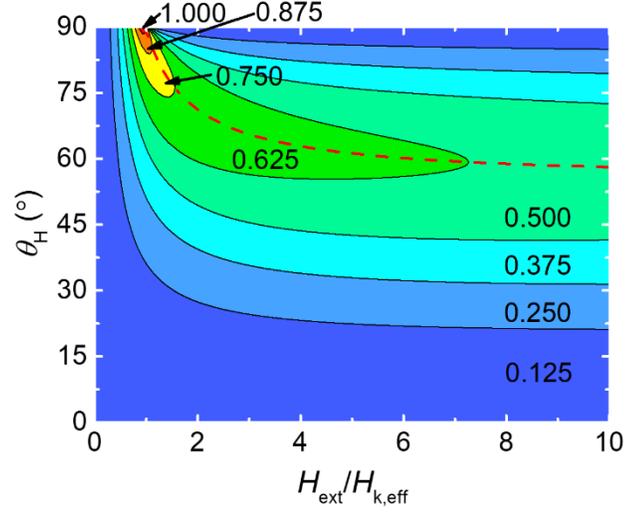

**Figure 3**. A contour plot of the relative signal size as a function of field ratio ($H_{ext}/H_{k,eff}$) and $\theta_H$ where a value of "1" indicates the maximum possible signal. The dotted line shows the $\theta_H$ where the signal is maximized at a specific field ratio.

For field-swept measurements, (where the angle is held constant and the field is swept) Fig. 4 should provide a simple guide for maximizing signals (a summary of $\theta_{H,MAX}$ in Fig. 3). To further assist in the design of TR-MOKE signals to maximize SNR, we suggest a simplified estimation for the determination of the amplitude of TR-MOKE signal. Equation 5 predicts the precession amplitude based on the equilibrium direction ($\theta$, from Fig. 1) and the external field angle. The magnitude of $H_{ext}$ is integrated into Eq. 5 through the $\theta$ through Eq. 6 which provides the minimum energy condition.

$$\frac{\Delta M_z}{M_s} \propto \sin(\theta)\sin(\theta - \theta_H) \qquad (5)$$

$$2H_{ext}\sin(\theta_H - \theta) = H_{k,eff}\sin(2\theta) \qquad (6)$$

This simplified expression is based on the product of the two components for signal maximization previously discussed: the projection of the magnetization in the z-direction, $\sin(\theta)$, and the magnetic torque, $\sin(\theta - \theta_H)$. While the simplified expression presented in Eq. 2 cannot capture all the details of a more complex LLG simulation, it is more than accurate enough for an initial estimate of $\theta_{H,MAX}$, as shown by the comparison in Fig. 4.

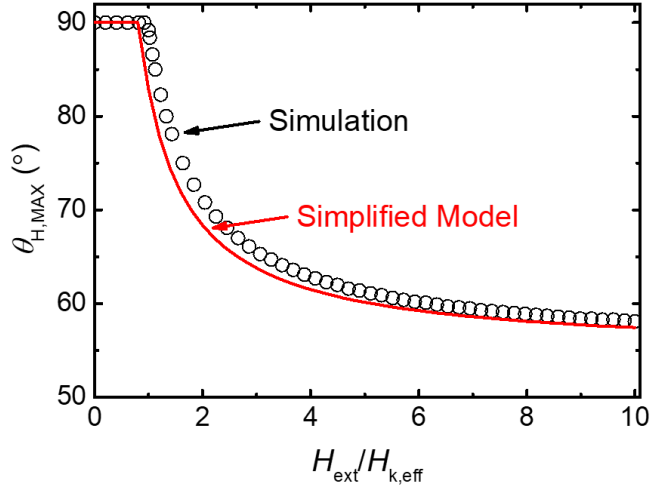

**Figure 4**. The trend of $\theta_{H,MAX}$ at a given field ratio. The open circles indicate results from the LLG simulation discussed in Section I, while the red curve is the simplified model from Eq. 5.

## III. COMPARING SIMULATION RESULTS TO TR-MOKE MEASUREMENTS

To verify the precited results for the maximum TR-MOKE signal amplitude, a series of measurements were conducted on a 300 °C post-annealed W/CoFeB/MgO film (see our previous publication for more information). After conducting measurements, the thermal background was subtracted leaving purely the decaying sinusoidal term. The oscillation amplitude from measurement was calculated as shown in Fig. 2a. Results from four values of $H_{ext}$ and six values of $\theta_H$ are summarized in Fig. 5.

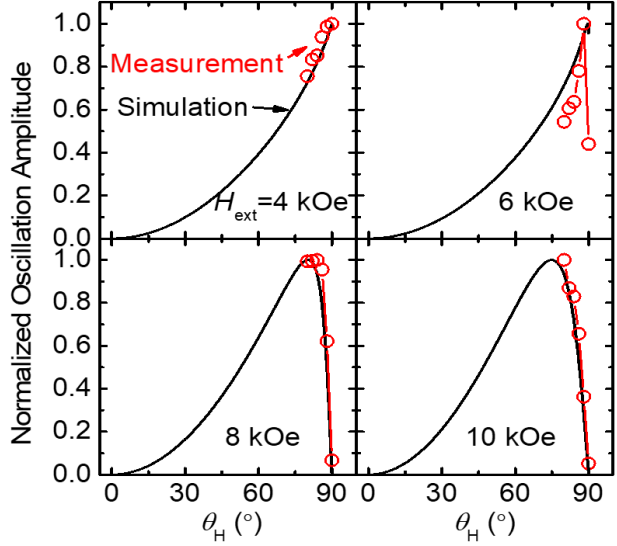

**Figure 5**. Normalized TR-MOKE oscillation amplitudes directly for a W/CoFeB/MgO when $H_{ext}$ is 4, 6, 8, and 10 kOe. The open red circles show the measurement data (a line between points is provided to guide the eye) while the black curves indicate the results from the LLG simulations for a material with $H_{k,eff} \approx 6$ kOe.

Comparisons between the trends predicted simulations and measurement results show remarkable agreement. As expected, the signal amplitude decreases with increasing angle for $H_{ext} < H_{k,eff}$ ($H_{k,eff} \approx 6$ kOe) and decreases with increasing angle for $H_{ext} > H_{k,eff}$. These measurements can even capture the predicted peak of amplitude at nearly the same $\theta_H$ for fields near $H_{k,eff}$. For the 6 kOe measurements, there is a slight deviation in the amount of decay in signal strength for decreasing $\theta_H$ (simulations predict a slower decrease). This is most likely due to an inhomogeneous broadening effect (i.e. the $H_{k,eff}$ in the sample has a distribution of values) leading to a deviation from theory near $H_{k,eff}$. While the $\theta_H$ in the setup used in this experiment was limited, these results verify that the excellent agreement between simulation and measurement.

## IV. CONCLUSION

In conclusion, we utilized a numerical approach to calculate the dynamic response of magnetization to a demagnetization process. We find that the size of the magnetic precession, and thus the size of the TR-MOKE signal depends on the angle and amplitude of the external field (relative to $H_{k,eff}$). To verify the results of these simulations, we conducted measurements on a W/CoFeB/MgO sample with perpendicular magnetic anisotropy. The results of the measurements show that the magnitude of the TR-MOKE signal shows good agreement with our prediction. These results should assist to maximize the SNR in TR-MOKE measurements.

## ACKNOWLEDGEMENTS

This work is supported by C-SPIN (award #: 2013-MA-2381), one of six centers of STARnet, a Semiconductor Research Corporation program, sponsored by MARCO and DARPA.

## REFERENCES

[1] Iida, S., 1963, "The difference between gilbert's and landau-lifshitz's equations," Journal of Physics and Chemistry of Solids, 24(5), pp. 625-630.
[2] van Kampen, M., Jozsa, C., Kohlhepp, J. T., LeClair, P., Lagae, L., de Jonge, W. J. M., and Koopmans, B., 2002, "All-Optical Probe of Coherent Spin Waves," Physical Review Letters, 88(22), p. 227201.
[3] Zhu, J., Wu, X., Lattery, D. M., Zheng, W., and Wang, X., 2017, "The Ultrafast Laser Pump-Probe Technique for Thermal Characterization of Materials With Micro/Nanostructures," Nanoscale and Microscale Thermophysical Engineering, 21(3), pp. 177-198.
[4] You, C.-Y., and Shin, S.-C., 1998, "Generalized analytic formulae for magneto-optical Kerr effects," Journal of Applied Physics, 84(1), pp. 541-546.